\documentclass[aps,prd,10pt,a4paper,superscriptaddress,preprintnumbers,showpacs,notitlepage,eqsecnum,nofootinbib]{revtex4-1}

\usepackage[english]{babel}
\usepackage{graphicx}
\usepackage{amsmath}

\def\sqr#1#2{{\vcenter{\vbox{\hrule height.#2pt\hbox{\vrule
width.#2pt height#1pt \kern#1pt\vrule width.#2pt}\hrule height.#2pt}}}}

\begin{document}

\title{ Disformal Transformations,\\
Veiled General Relativity and Mimetic Gravity\\
}

\author{Nathalie Deruelle}

\affiliation{
APC, CNRS-Universit\'e  Paris 7, 75205 Paris CEDEX 13, France.}

\author{Josephine Rua}
\affiliation{Instituto de Cosmologia, Relatividade e Astrofísica - ICRA/CBPF, Rua Dr. Xavier Sigaud 150 - 22290-180 Rio de Janeiro - Brazil}

\begin{abstract}

In this Note we show that Einstein's equations for gravity are generically invariant under ``disformations". We also show that the particular subclass when this is not true yields the equations of motion of ``Mimetic Gravity". Finally we give the ``mimetic" generalization of the Schwarzschild solution.

\end{abstract}

\maketitle

\section{Introduction}

In scalar-tensor theories where gravity is described  by a metric $g_{\mu\nu}$ together with an extra scalar field $\Psi$ one can work either in the ``Jordan frame" where the action for gravity is not the Einstein-Hilbert action for $g_{\mu\nu}$ (e.g. $f(R)$ instead of $R$) but where matter is minimally coupled to  $g_{\mu\nu}$, or in the ``Einstein frame" where the action for gravity is Einstein-Hilbert's for a conformally related metric $\ell_{\mu\nu}=g_{\mu\nu}/F(\Psi(x^\mu))$ but where matter is non-minimally coupled to $\ell_{\mu\nu}$. Working in one or the other ``frame" is a matter of taste and the observable predictions are the same (up to pathological cases when the conformal factor $F$ diverges or goes through zero).

The equivalence of the two descriptions is particularly striking when one considers strict General Relativity, where the action for gravity is Einstein-Hilbert's  with matter minimally coupled to the metric, see \cite{DerSas}. Indeed, if one rewrites the action in terms of the metric $\ell_{\mu\nu}=g_{\mu\nu}/F(\Psi(x^\mu))$ so that it looks like the action for a scalar-tensor theory, the equations of motion obtained by extremizing the action with respect to $\ell_{\mu\nu}$ and $\Psi$ will look quite different from the Einstein equations but, of course, can be transformed back to the Einstein equations for $g_{\mu\nu}$ by performing the inverse transformation, $\ell_{\mu\nu}=F g_{\mu\nu}$, hence the name ``veiled" General Relativity (or, rather ``Weyled" GR~!).

Now,  on one hand, transformations more general than conformal ones have long been studied, in particular the ``disformations"  first considered by Bekenstein \cite{Bek}, such that
 \begin{equation}
 g_{\mu\nu}=F(\Psi,w)\,\ell_{\mu\nu}+H(\Psi,w)\,\partial_\mu\Psi\partial_\nu\Psi\,,\quad\hbox{where}\quad w\equiv\ell^{\rho\sigma}\partial_\rho\Psi\partial_\sigma\Psi\label{disformed metric intro}
 \end{equation}
 and where the functions $F$ and $H$ are a priori arbitrary. The invariance, under such transformations, of field equations such as the Klein-Gordon, the Maxwell or the Horndeski equations of motion has recently been studied in, respectively,  \cite{GouFalKG}, \cite{GouFalMax}, \cite{Zum} - \cite{BetLib} and references therein.
 
 On the other hand, a particular subclass of disformed metrics with $F=w$ and $H=0$, to wit
 \begin{equation}
 g_{\mu\nu}=w\,\ell_{\mu\nu}\,,\label{mimetic metric intro}
 \end{equation}
 was considered in \cite{ChaMuk} where the action for standard General Relativity with matter minimally coupled to $g_{\mu\nu}$ was extremized, not with respect to $g_{\mu\nu}$, but with respect to $\ell_{\mu\nu}$ and $\Psi$. Somewhat surprinsigly it was found that the equations of motion where not just the ``veiled" Einstein equations but included extra branches of solutions. Cosmological solutions of the equations of motion for this ``Mimetic Gravity" were studied in \cite{ChaMuk} where it was shown that they can mimic the contribution of an extra pressure-less fluid such as Dark Matter. Comments on their structure were made in \cite{Bar}. In  \cite{Gol}  it was shown that they can also be deduced by extremizing, with respect to $g_{\mu\nu}$, the action $S+\int\,\lambda\sqrt{-g}(g^{\mu\nu}\partial_\mu\Psi\partial_\nu\Psi-1)$ where $S$ is the Einstein-Hilbert action  and where $\lambda$ is a Lagrange multiplier. This result was exploited  in \cite{ChaMukbis} where the cosmological model  \cite{ChaMuk} was extended to include a description of Dark Energy as well. Finally Hamiltonian formulations were presented in \cite{ChaKlu} and \cite{Mal}.

The purpose of this  Note is first to show (in Section \ref{veiled GR}) that, in accordance with the standard expectation inherited from the conformal equivalence of Jordan and Einstein frames in scalar-tensor theories of gravity, the equations of motion deduced from the action for standard General Relativity with matter minimally coupled to $g_{\mu\nu}$ are the Einstein equations, whether one extremizes it as usual with respect to the metric $g_{\mu\nu}$, or with respect to $\ell_{\mu\nu}$ and $\Psi$, when both are related by the {\it generic} disformation (\ref{disformed metric intro}).

We will then see (in Section  \ref{mimetic grav}) that there is however a particular class of disformations where the function $H$ is given by
\begin{equation}
H(w,\Psi)= -{F(w,\Psi)\over w}+h(\Psi)\label{mimetic H intro}\,,
\end{equation}
to which the metrics (\ref{mimetic metric intro}) belong, for which the equations of motion, first, are not Einstein's and, second, turn out to reduce to the ``Mimetic" equations of motion obtained in \cite{ChaMuk} when the vector $\partial_\mu\Psi$ is timelike.

\vfill\eject

Finally (in Section \ref{static Sphe}) we give, for  spacelike $\partial_\mu\Psi$, the non trivial (that is non-Schwarzschild)  vacuum, static and spherically symmetric  solution of Mimetric Gravity and find that 
it widly differs from the well-tested Schwarzschild solution. We conclude in Section \ref{conclusion} with a few comments for further work.

\section{Equations of motion}

Consider the standard Einstein-Hilbert action
\begin{equation}
S={1\over2\kappa}\int R\sqrt{-g}\,d^4x+S_{\rm m}[\phi_{\rm m},g_{\mu\nu}]\,.\label{EH action}
\end{equation}
 $\kappa\equiv8\pi G$ is Einstein's constant,  $g$ is the determinant of the components  $g_{\mu\nu}$  of a metric in some coordinates $x^\mu$ (signature $-+++$) and $R$ is its scalar curvature~; the matter fields $\phi_{\rm m}$ are minimally coupled to the metric.
 
 In General Relativity the metric $g_{\mu\nu}$ describes gravity  and  Einstein's equations  are obtained by extremizing the action with respect to its variations. Here, the fundamental fields describing gravity are taken to be another metric $\ell_{\mu\nu}$ and a scalar field $\Psi$ which define $g_{\mu\nu}$  by means of a ``disformation"\cite{Bek}~:
 \begin{equation}
 g_{\mu\nu}=F(\Psi,w)\,\ell_{\mu\nu}+H(\Psi,w)\,\partial_\mu\Psi\partial_\nu\Psi\quad\hbox{where}\quad w\equiv\ell^{\rho\sigma}\partial_\rho\Psi\partial_\sigma\Psi\,.\label{disformed metric}
 \end{equation}
 The functions $F$ and $H$ are a priori arbitrary.\footnote{We shall assume $F\neq0$.} In ``Mimetic Gravity" \cite{ChaMuk}, $F=w$ and $H=0$.

Varying the action with respect to $\ell_{\mu\nu}$ and $\Psi$ is straightforward. 

First we have, up to boundary terms that we shall ignore,
\begin{equation}
\delta S=-{1\over2\kappa}\int\!d^4x\sqrt{-g}(G^{\mu\nu}-\kappa T^{\mu\nu})\delta g_{\mu\nu}\quad\hbox{where}\quad T^{\mu\nu}\equiv{2\over\sqrt{-g}}{\delta S_{\rm m}\over\delta g_{\mu\nu}}
\end{equation}
and where $G_{\mu\nu}$ is Einstein's tensor.
Invariance of the action under diffeomorphisms $x^\mu\to x^\mu+\xi^\mu\ \Longrightarrow\ \delta g_{\mu\nu}=D_\mu\xi_\nu+D_\nu\xi_\mu$, together with Bianchi's identity $D_\nu G^\nu_\mu\equiv0$,   guarantees that the matter energy-momentum tensor  $T^{\mu\nu}$ is conserved,
\begin{equation}
D_\nu T^\nu_\mu=0\,,
\end{equation}
where $D_\mu$ is the Levi-Civita covariant derivative associated with the metric $g_{\mu\nu}$ and where indices are moved with $g_{\mu\nu}$~: $T^\nu_\mu\equiv g_{\mu\rho}T^{\nu\rho}$.

Second, we have
\begin{equation}
\begin{aligned}
\delta g_{\mu\nu}&=F\,\delta\ell_{\mu\nu}-\left(\ell_{\mu\nu}{\partial F\over\partial w}+\partial_\mu\Psi\partial_\nu\Psi{\partial H\over \partial w}\right)\left[(\ell^{\alpha\rho}\partial_\alpha\Psi)\,
(\ell^{\beta\sigma}\partial_\sigma\Psi)\,\delta\ell_{\rho\sigma}-2\ell^{\rho\sigma}(\partial_\rho\Psi)\,(\partial_\sigma\delta\Psi)\right]\\
&+\left(\ell_{\mu\nu}{\partial F\over\partial  \Psi}+\partial_\mu\Psi\partial_\nu\Psi{\partial H\over \partial \Psi}\right)\delta\Psi +H\left[(\partial_\mu\Psi)(\partial_\nu\delta\Psi)+(\partial_\nu\Psi)(\partial_\mu\delta\Psi)\right]\,.
\end{aligned}
\end{equation}

Hence the following equations of motion~: $\delta S/\delta\ell_{\mu\nu}=0$, that is,
\begin{equation}
\begin{aligned}
F(G^{\mu\nu}-\kappa T^{\mu\nu})&=\left(A{\partial F\over\partial w}+B{\partial H\over\partial w}\right)(\ell^{\mu\rho}\partial_\alpha\Psi)\,
(\ell^{\nu\sigma}\partial_\sigma\Psi)\\
\quad\hbox{where}\quad  A&\equiv (G^{\rho\sigma}-\kappa T^{\rho\sigma})\ell_{\rho\sigma}\quad\hbox{and}\quad B\equiv(G^{\rho\sigma}-\kappa T^{\rho\sigma})\partial_\rho\Psi\,\partial_\sigma\Psi\,,
\end{aligned}
\label{ell eom}
\end{equation}
and $\delta S/\delta\Psi=0$, that is,
\begin{equation}
{2\over\sqrt{-g}}\partial_\rho\left\{\sqrt{-g}\,\partial_\sigma\Psi\left[H(G^{\rho\sigma}-\kappa T^{\rho\sigma})+\left(A{\partial F\over\partial w}+B{\partial H\over\partial w}\right)\ell^{\rho\sigma}\right]
\right\}=
A{\partial F\over\partial\Psi}+B{\partial H\over\partial\Psi}\,.\label{psi eom}
\end{equation}
(Note that in theories such as ``Geometric Scalar gravity" the metric $\ell_{\mu\nu}$ is a background, flat, metric and is kept fixed so that the equations of motion reduce to an equation for the scalar field $\Psi$ only, such as (\ref{psi eom}), see \cite{GSG}.)

\section{The generic case~: Veiled General Relativity\label{veiled GR}}

Contractions of the equation of motion (\ref{ell eom}) with $\ell_{\mu\nu}$  and  $\partial_\mu\Psi\,\partial_\nu\Psi$ yield
\begin{equation}
A\left(F-w{\partial F\over\partial w}\right)-Bw{\partial H\over\partial w}=0\quad,\quad A\,w^2{\partial F\over\partial w}-B\left(F-w^2{\partial H\over\partial w}\right)=0\,.\label{det eq}
\end{equation}
The determinant $det$ of the system is
\begin{equation}
det=w^2F{\partial\ \over\partial w}\left(H+{F\over w}\right)\,.\label{det}
\end{equation}
In the generic case when it is not zero the only solution of (\ref{det eq}) is $A=B=0$ and the equations of motion (\ref{ell eom}) and (\ref{psi eom}) reduce to
\begin{equation}
F(G^{\mu\nu}-\kappa T^{\mu\nu})=0\quad,\quad \partial_\rho\left[\sqrt{-g}\,\partial_\sigma\Psi H(G^{\rho\sigma}-\kappa T^{\rho\sigma})
\right]=0
\end{equation}
where $F\neq0$. Hence the first equation is Einstein's equations and the second is empty. 

In the generic case then, the extremizations of the Einstein-Hilbert action (\ref{EH action}) with respect to the ``disformed" metric $g_{\mu\nu}$ or with respect to its elements $\ell_{\mu\nu}$ and $\Psi$ are equivalent and yield the standard Einstein equations of General Relativity $G_{\mu\nu}=\kappa T_{\mu\nu}$ for the metric $g_{\mu\nu}$. The theory is then nothing more than a generalization of ``veiled" General Relativity where the disformed metric $g_{\mu\nu}$ reduces to a conformal metric $g_{\mu\nu}=F(\Psi)\ell_{\mu\nu}$, see e.g. \cite{DerSas}.

\section{Mimetic gravity\label{mimetic grav}}

Let us now turn to the case when the determinant (\ref{det}) is zero. Since $F\neq0$ this implies that the function $H(w,\Psi)$ is constrained to be of the form
\begin{equation}
H(w,\Psi)= -{F(w,\Psi)\over w}+h(\Psi)\label{mimetic H}\,.
\end{equation}
(Note that ``Mimetic Gravity" where $F(w,\Psi)$ and $h(\Psi)=1$ falls in that class.)
The solution of the system (\ref{det eq}) then is $B=w\,A$ and the equations of motion (\ref{ell eom}) and (\ref{psi eom})  simplify into
\begin{equation}
G^{\mu\nu}-\kappa T^{\mu\nu}={A\over w}\,(\ell^{\mu\rho}\partial_\rho\Psi)\,
(\ell^{\nu\sigma}\partial_\sigma\Psi)\quad,\quad
{2\over\sqrt{-g}}\partial_\rho\left(\sqrt{-g}\,h\,A\,\ell^{\rho\sigma}\partial_\sigma\Psi\right)=A\,w{dh\over d\Psi}\label{mixed eom}
\end{equation}
where we recall that $A\equiv (G^{\rho\sigma}-\kappa T^{\rho\sigma})\ell_{\rho\sigma}$.

When $h(\Psi)\neq0$, which we shall suppose, these equations of motion can be written in terms of the disformed metric $g_{\mu\nu}$ only. Indeed we have from (\ref{disformed metric}) and (\ref{mimetic H})~:
\begin{equation}
g_{\mu\nu}=F(\Psi,w)\,\ell_{\mu\nu}+\partial_\mu\Psi\,\partial_\nu\Psi\left(h(\Psi)-{F(\Psi,w)\over w}\right)\,,
\end{equation}
which is invertible if $h\neq0$, with
\begin{equation}
g^{\mu\nu}={\ell^{\mu\nu}\over F}+{F-w\,h\over F\,h\,w^2}(\ell^{\mu\rho}\partial_\rho\Psi)\,
(\ell^{\nu\sigma}\partial_\sigma\Psi)\,.
\end{equation}
Hence we have that $A=(G-\kappa T)/(hw)$ and $\ell^{\mu\rho}\partial_\rho\Psi=h\,w\,\partial^\mu\Psi$, where $G-\kappa T\equiv g_{\rho\sigma}(G^{\rho\sigma}-\kappa T^{\rho\sigma})$ and $\partial^\mu\Psi\equiv g^{\mu\rho}\partial_\rho\Psi$.  Therefore the equations of motion (\ref{mixed eom}) read
\begin{equation}
G_{\mu\nu}-\kappa T_{\mu\nu}=(G-\kappa T)\,h\,\partial_\mu\Psi\,\partial_\nu\Psi\quad,\quad
2D_\rho\left[(G-\kappa T)h\,\partial^\rho\Psi\right]=(G-\kappa T) {1\over h}{dh\over d\Psi}\,.
\end{equation}
Finally the function $h(\Psi)$ can be eliminated by means of a field redefinition. Indeed, introducing $\Phi$ such that $d\Phi/d\Psi=\sqrt{|h|}$ we finally get
\begin{equation}
G_{\mu\nu}-\kappa T_{\mu\nu}=\epsilon(G-\kappa T)\partial_\mu\Phi\,\partial_\nu\Phi\quad,\quad
2D_\rho\left[(G-\kappa T)\partial^\rho\Phi\right]=0\,,\label{mimetic eom}
\end{equation}
where $\epsilon=\pm1$ depending on the sign of the norm of $\partial_\mu\Psi$~: $g^{\mu\nu}\partial_\mu\,\Phi\partial_\nu\Phi=\epsilon$.

For $\epsilon=-1$, that is, for timelike $\partial_\mu\Phi$, these equations are exactly the same as the original equations of motion for mimetic gravity which were derived in  \cite{ChaMuk} in the particular case when $h(\Psi)=1$ and $F(\Psi,w)=w$.

\vfill\eject

What was done here was to show that they are also obtained by varying the Einstein-Hilbert action (\ref{EH action}) with respect to the fields $\ell_{\mu\nu}$ and $\Phi$ which define a more general disformed metric $g_{\mu\nu}$ given by
\begin{equation}
g_{\mu\nu}=F(\Psi,w)\,\ell_{\mu\nu}\pm\,\partial_\mu\Phi\,\partial_\nu\Phi\left(1-{F(\Psi,w)\over w\,h(\Psi)}\right)
\end{equation}
where $F(\Psi,w)$ and $h(\Psi)$ are arbitrary and where $g^{\mu\nu}\partial_\mu\Phi\,\partial_\nu\Phi=\pm1$. Moreover we saw in the previous section that
if the disformed metric is not of the above type then the equations of motion are simply the Einstein equations, $G_{\mu\nu}=\kappa\,T_{\mu\nu}$.

\section{Static spherically symmetric solutions in vacuum\label{static Sphe}}

As shown in \cite{ChaMuk}, equations (\ref{mimetic eom}) have no vacuum, spherically symmetric, static, solution if $\partial_\mu\Phi$ is timelike.

Here we give their vacuum, static, spherically symmetric solution in the spacelike sector of $\partial_\mu\Phi$ . 

Schwarzschild-Droste coordinates $x^\mu=(t,r,\theta,\phi)$ are chosen so that the field $\Phi$ and the line element $ds^2\equiv g_{\mu\nu}dx^\mu\,dx^\nu$  are manifestly static and spherically symmetric~:
\begin{equation}
ds^2=-e^{\nu(r)}dt^2+e^{\lambda(r)}dr^2+r^2(d\theta^2+\sin^2\theta\,d\phi^2)\quad\hbox{and}\quad \Phi=\Phi(r)\,.
\end{equation}
First one notes that the Schwarzschild metric, which is such that $G_{\mu\nu}=0$  solves  equation (\ref{mimetic eom}), for all functions $\Phi(r)$. 

When now the scalar curvature $R=-G$ is not zero,
the trace of the first equation (\ref{mimetic eom}) imposes to choose $\epsilon=+1$  and $\partial^\mu\Phi\,\partial_\mu\Phi=1$, that is~: $d\Phi/dr=e^{\lambda/2}$ (therefore $\Phi$ is the radial gaussian normal coordinate). The second equation then gives ~: $R=C_0e^{-\nu/2}/r^2$. 
Recalling that
\begin{equation}
G_{tt}={e^\nu\over r^2}{d\ \over dr}\left[r\left(1-e^{-\lambda}\right)\right]\quad\hbox{and}\quad G_{rr}=-{e^\lambda\over r^2}\left(1-e^{-\lambda}\right)+{1\over r}{d\nu \over dr}
\end{equation}
the equations $G_{tt}=0$ and $G_{rr}=-R(d\Phi/dr)^2$ then solve as
\begin{equation}
e^{-\lambda}=1-2m/r\quad,\quad e^{\nu/2}=C_1\sqrt{1-2m/r}+C_0\left[1-\sqrt{1-2m/r}\ln\left(\sqrt r+\sqrt{r-2m}\right)\right]
\end{equation}
and one can check that the other equation, $G_{\theta\theta}=0$, is indeed satisfied (thanks to the Bianchi identity). The constant $C_1$ can be absorbed in a redefinition of the time coordinate $t$ and hence can be set equal to 1. The solution therefore depends of two constants, $m$ and $C_0$. When $C_0=0$ it reduces to the standard Schwarzschild solution.

For $C_0\neq0$ the metric is not asymptotically flat and differs strongly from the well tested Schwarzschild metric.

This is an indication that ``Mimetic Gravity" may provide an explanation for the presence of Dark Matter in the universe in the sector where $\partial_\mu\Phi$ is timelike, but is at odds with tests in the solar system in the sector where $\partial_\mu\Phi$ is spacelike.

\section{Conclusion\label{conclusion}}

In this Note we showed that General Relativity is generically invariant under field redefinitions given by means of disformations. We also showed that the particular sub-class when this is not the case reduces to Mimetic Gravity. Finally we saw that the general vacuum static spherically symmetric solution of Mimetic Gravity in the sector where $\partial_\mu\Phi$ is spacelike is pathological.

A important feature of ``Mimetic Gravity" is that  the field $\Phi$, whose gradient is constrained to be of unit norm, is not dynamical. Rendering it dynamical may provide ways to improve the theory. We leave this perspective to further work.

\begin{acknowledgments}

The authors thank V. Mukhanov and A. Vikman for very useful correspondence.

N.D. is grateful to the hospitality of CBPF in December 2013.

J.R. acknowledges financial support from ``Coordena\c c$\tilde{\rm a}$o de Aperfeiçoamento de Pessoal de N\` ivel Superior" (CAPES) of Brazil and the ``Programme Antenne Br\'esil" run by the "Service des Relations Internationales et Europ\'eennes de l'Universit\'e Sorbonne Nouvelle - Paris 3" and thanks APC for its hospitality.

\end{acknowledgments}

\vfill\eject

\end{document}